\documentclass[a4paper,fleqn,usenatbib]{mnras}

\usepackage{newtxtext,newtxmath}

\usepackage[T1]{fontenc}
\usepackage{ae,aecompl}
\usepackage{soul}
\soulregister\cite7 
\soulregister\citep7 
\soulregister\citet7 
\soulregister\ref7 
\soulregister\pageref7 

\usepackage{graphicx}	
\usepackage{amsmath}	
\usepackage{amssymb}	

\title[Characterization of precision premium]{Characterization of precision premium in astrometry}

\author[F. R. Lin et al.]{
F. R. Lin,$^{1,3}$
J. H. Peng,$^{2,3}$
Z. J. Zheng$^{1,3}$ and
Q. Y. Peng$^{1,3}$\thanks{tpengqy@jnu.edu.cn}
\\
$^{1}$Department of Computer Science, Jinan University, Guangzhou 510632, China\\
$^{2}$Department of Physics and Astronomy, University of Victoria, Elliott Building, 3800 Finnerty Rd., Victoria, BC V8P 5C2, Canada\\
$^{3}$Sino-French Joint Laboratory for Astrometry, Dynamics and Space Science, Jinan University, Guangzhou 510632, China
}

\pubyear{2019}

\begin{document}
\label{firstpage}
\pagerange{\pageref{firstpage}--\pageref{lastpage}}
\maketitle

\begin{abstract}
Precision premium, a concept in astrometry that was firstly presented by Pascu in 1994, initially means that the relative positional measurement of the Galilean satellites of Jupiter would be more accurate when their separations are small. Correspondingly, many observations have been obtained of these Galilean satellites since then. However, the exact range of the separation in which precision premium takes effect is not clear yet, not to say the variation of the precision with the separation.
In this paper, the observations of open cluster M35 are used to study precision premium and the newest star catalogue Gaia DR2 is used in the data reduction. Our results show that precision premium does work in about less than 100 arcsecs for two concerned objects, and the relative positional precision can be well fitted by a sigmoidal function.
Observations of Uranian satellites are also reduced as an example of precision premium.
\end{abstract}

\begin{keywords}
astrometry - methods: analytical - methods: statistical
\end{keywords}



\section{Introduction}

The astrometry of natural satellites is a fundamental work in the study of their orbital theory, which is very important in the research of formation and evolution of the solar system. For some satellites, obtaining a precise measurement from differential CCD astrometry is not easy because of the lack of reference stars. A good method to obtain high-precision astrometric data of these satellites is to observe the light curves of mutual events of them \citep[to name a few]{Arlot2012, Arlot2013, Assafin2009, Ziese2018}. So the mutual events of the satellites have attracted great attention, theoretical predictions \citep{Arlot2006, Arlot2008} have been done to facilitate observations to be taken all over the world. The observations of mutual events have provided many valuable data for the study of natural satellites \citep{Lainey2009, Arlot2019}. However, this type of observation is rare, since the mutual event happens only when the Earth and the Sun pass through the orbital planes of the satellites. The occurring period of this phenomenon is about 6 years for Galilean satellites, 15 years for Saturnian satellites and as long as 42 years for the satellites of Uranus \citep{Arlot2012, Arlot2013, Arlot2014}.

Another effective technique to obtain high-precision astrometric data of these satellites is to observe them when their separation is small enough \citep{Peng2008}. This technique is based on precision premium, which was firstly presented by \citet{Pascu1994} and was used to deliver the observations of the Galilean satellites.
Using this technique, many high-precision astrometric results of Galilean satellites have been obtained \citep[e.g.][]{Pascu1996, Peng2008, Peng2012a}. Specifically, \citet{Pascu1994} chose observations with a small separation of 50 arcsecs (and below) to obtain the high-precision results with an external precision of $\pm$0.01 arcsec. \citet{Peng2012a} focused on the relative position of satellite pairs with their separation less than 85 arcsecs and their results showed that the precision was better than 30 mas. One advantage of the precision premium observations is that the satellites are close enough every few days, so it is relatively frequent to observe them. A recently-developed technique called mutual approximation was made by \citet{Morgado2016}, who delivered the instant (and other measurements) for two concerned satellites when the satellites have their minimum apparent separation in the sky plane. The mutual approximation technique chose arbitrarily the approximations for which the separation of satellite pair is smaller than 30 arcsecs. \citet{Morgado2019} successfully used this technique to the Galilean satellites and obtained an average internal precision of 11.3 mas. The extremely high precision and practicability of this technique indicate that it has great potential in astrometry for some bright moving objects. Moreover, the precision premium provides inter-object positions. For Galilean (or Uranian) satellites those are inter-satellite positions and not absolute values, in other words, these relative positions are not affected by the position of Jupiter (or Uranus).

Although precision premium has been used in the observation of the Galilean satellites for about one decade, the relation between precision improvement and the separation of the satellites is not clear yet. Therefore, the choice of the separation mentioned above is obviously arbitrary. A detailed characterization of precision premium is necessary to make widely use of it in astrometric measurements. As such, it's essential to obtain a large amount of relative position measurements with different separations. However, obtaining so many measurements of the Galilean satellites in a short period of time is not easy. We have ever tried to use observations of a bright asteroid passing through some dense star field to derive the relation between separation and its precision, but the measurements are found to be still not enough. Considering that the effect of precision premium should be independent of whether an object moves with respect to its reference object, a better choice is to use observations of an open cluster in which hundreds of the objects with different separations can be measured.

In this paper, observations of the open cluster M35 captured in November 2018 taken with the 1 m telescope at Yunnan Observatory are reduced by a classical reduction procedure. Then a simple but effective statistical method is used to analyze the reduction results. A sigmoidal function is found to be suitable to fit the statistics well and the function is referred to as precision premium curve (PPC). Our statistics show that precision premium is significant for the observations of the open cluster. The improvement of the precision due to precision premium increases almost linearly with the decrease of the objects' separation in a certain range. This work demonstrates that precision premium generally exists in conventional observations.
That is to say, the relative techniques for position measurement of Galilean satellites, such as the mutual approximation technique, can be adapted to other objects bright enough.
As an example, we reduce the observations of the Uranian satellites, which are bright enough, to show the universality of precision premium and the possibility to reach a high precision of several \emph{mas} for the bright objects.

The contents of this paper are arranged as follows. In Section 2, the observations used in this paper and corresponding instruments used to capture them are presented in detail. The method to generate the PPC and analysis of its characteristics are given in Section 3, as well as an example of taking advantage of precision premium to improve the reduction results.
Finally, some conclusions are drawn in Section 4.


\section{Observations and instruments}
From November 11 to 13, 2018, three-night observations of the open cluster M35 and the Uranian satellites (only on November 11) were captured by the 1 m telescope at the Yunnan Observatory (IAU code 286, longitude\---E$102^{\circ}47^{\prime}18^{\prime\prime}$, latitude--N$25^{\circ}1^{\prime}30^{\prime\prime}$, and height--2000 m above sea level). The exposure time of each frame is 60.0 s for M35 and 120.0 s for Uranian satellites. Table~\ref{tab:observation} lists the detailed information of the observations used in this paper.
The specifications of the telescope and the CCD chip are presented in Table~\ref{tab:telescope}. Using a program of 2D Gaussian centering developed by us, more than 1000 stars can be searched in each frame of CCD image. For the CCD images of Uranian satellites, about 75 reference stars can be searched in each frame. The multiple exposures of the open cluster in each night were taken by the dithered observational scheme (``+'' type), which was presented in \citet{Peng2012b}. Therefore, many star pairs with different separations can be measured and used in our statistics.
\begin{table*}
	\centering
	\caption{Detailed information of the observations. The 2nd column is the number of frames in each observation set and the corresponding filter. The 3rd column is the zenith distance of the observations and the 4th column is the FWHM derived from a typical bright star by a 2D Gaussian fit. The 5th column lists the observational objects.}
	\label{tab:observation}
	\begin{tabular}{ccccc} 
		\hline
		Date & No. and Filter & ZD & FWHM (mean)& Object\\
         (y-m-d) &  & (degree) & (arcsec)&   \\
		\hline
		2018-11-11 & 55I & $27.4-6.9$& $1.4-2.3(1.77)$& M35\\
		2018-11-12& 48I & $18.1-4.5$& $1.7-2.2(1.88)$& M35 \\
		2018-11-13 & 52I & $17.2-0.7$&$1.5-2.0(1.66)$& M35\\
		2018-11-11 & 28I & $19.4-14.5$& $1.3-2.1(1.67)$& Uranian satellites\\
		\hline
	\end{tabular}
\end{table*}

\begin{table}
	\centering
	\caption{Specifications of the telescope and CCD chip}
	\label{tab:telescope}
	\begin{tabular}{lc} 
		\hline
      Focal length & 1330 cm\\
      $F$ ratio & 13\\
      Diameter of primary mirror & 101.6 cm\\
      Detector & Andor iKon-XL \\
      CCD field of view & $16' \times 16' $  \\
      Size of pixel & 15 $\mu \times 15  \mu$ \\
      Size of CCD array & $4096 \times 4112 $\\
      Angular extent per pixel & $\sim$0.234 arcsec pixel$^{-1}$\\
		\hline
	\end{tabular}
\end{table}

\section{Methods and Analysis}
Reduction of the open cluster observations is briefly described in this section. After we obtained the positional residual (observed minus computed; $O-C$) of each star, the statistical method used to generate the PPC is presented. Then characterization of precision premium and some suggestions of using it to improve the reduction method are given. An example of using precision premium in the reduction of Uranian satellites' observations is given at the end of this section.

\subsection{Data Reduction}\label{sec:reduction}

Each frame of CCD image is reduced by the following procedures. Firstly, a two-dimensional Gaussian centering algorithm is used to measure the pixel position ($x,y$) of each star. Then the standard coordinate ($\xi,\eta$) of each star can be calculated via the central projection \citep{Green1985}. Its reference equatorial coordinates ($\alpha,\delta$) here are taken from the newest star catalogue Gaia DR2 \citep{Gaia2018} and transformed to the observational epoch (i.e. topocentric apparent places, including atmosphere refraction). Finally, a weighted least squares scheme is used to solve the plate model with a fourth-order polynomial \citep{Peng2010}. The weight of each star in the least squares solution is designed by
\begin{equation}
w=1/\sigma^2(m),
\end{equation}
where $m$ is the magnitude of the star and $\sigma(m)$ is a sigmoidal function describing the relation between its magnitude and its measurement precision. $\sigma(m)$ can be expressed as
\begin{equation}
\sigma(m)=(A_1-A_2)/(1+e^{(m-m_0)/dm})+A_2,
\end{equation}
where $A_1 ,A_2, m_0$ and $dm$ are the free parameters. Fitting this function to the Mag-SD data (as shown in the left panels of Figure~\ref{fig:mpData}), the parameters for each observation set can be obtained (list in Table~\ref{tab:curveParam1}).

In this way, the $(O-C)$ of each star can be calculated. Then we have the information of <source ID, $\xi$, $\eta$, $(O-C)_\alpha$ ,$(O-C)_\delta$> for each star. The standard coordinates will be here used to calculate the separation between each star pair.

The effect of differential chromatic refraction (DCR) is ignored in the reduction because all CCD frames were taken when the zenith distance of each target field was small (see Table~\ref{tab:observation}). Besides, the Johnson $I$ filter used for observation can also reduce the effect of DCR. For all our observations, their zenith distances are smaller than 30 degrees, the corresponding DCR influences are no more than 3 mas according to \citet{Stone2002}.

\subsection{Statistics and analysis}\label{sec:analysis}

In order to minimize the systematic error between each star pair, the mean separation residual of each star pair should be subtracted first. Suppose there are $N$ stars observed in a dithered observation (each star may appear one or more times in each observation set), then $N(N-1)/2$ pairs of stars can be combined. If a star pair appear in less than 3 frames, they will be rejected from our statistics. Otherwise, a mean separation residual $M$ will be included in our statistics. Then the $M$ is subtracted from each separation residual of the star pair to obtain their real residual.

\begin{figure}
\begin{center}
	\includegraphics[width=0.9\columnwidth]{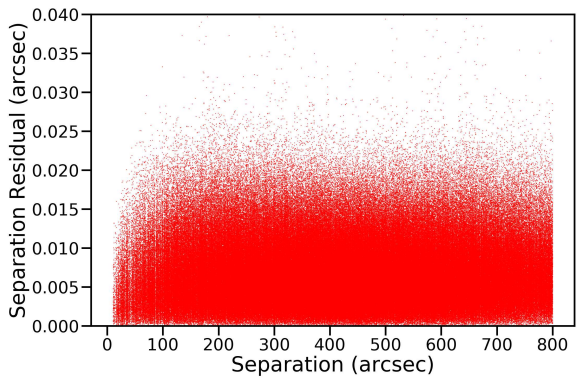}
    \caption{The separation and its residual of each star pair in each frame. The horizontal axis is the separations of the star pairs calculated using their standard coordinates. The vertical axis is their residuals. Stars fainter than 14 Gaia G-mag are not used in this process.}
    \label{fig:scatter}
\end{center}
\end{figure}

\begin{figure*}
    \begin{minipage}{0.45\textwidth}
	\includegraphics[width= \columnwidth]{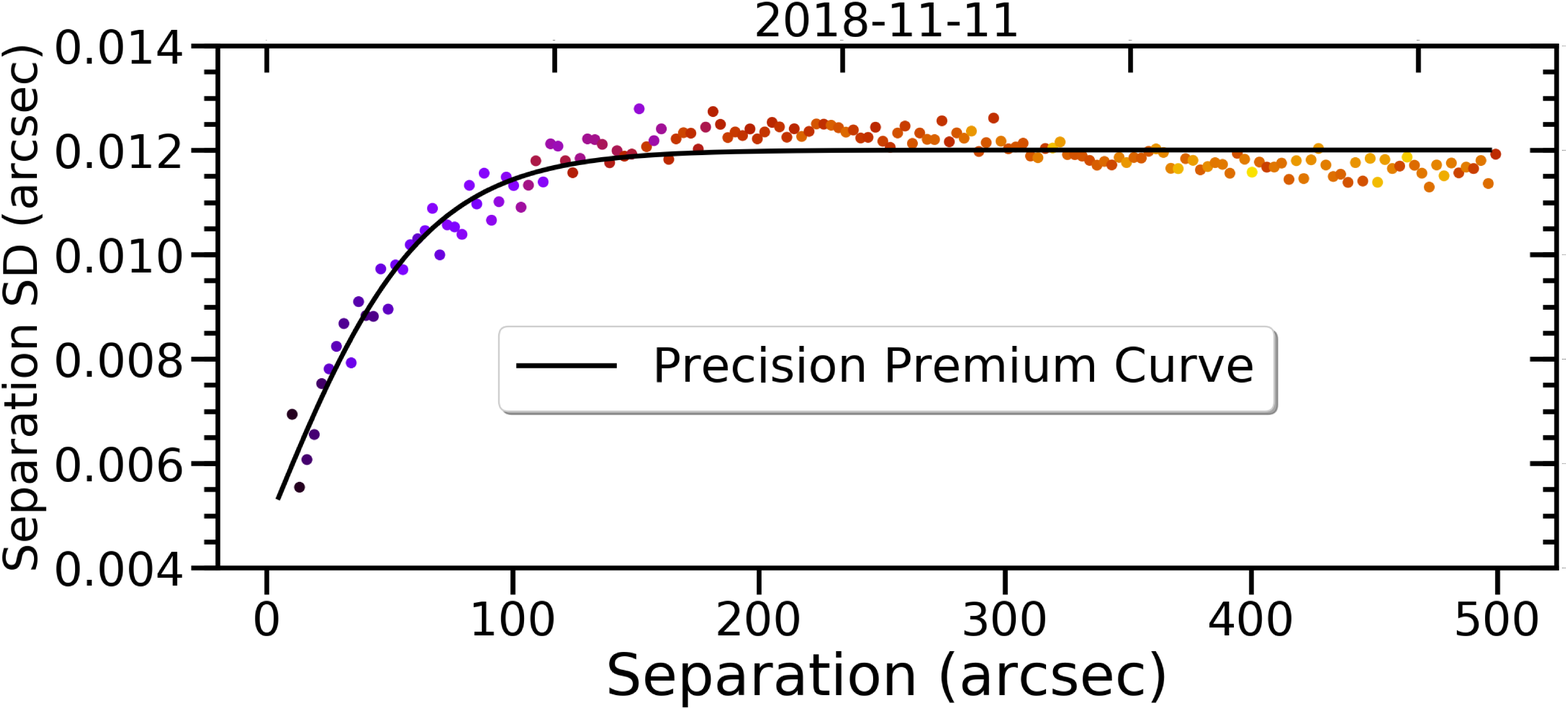}
    \end{minipage}
    \hspace{0.05in}
    \begin{minipage}{0.461\textwidth}
	\includegraphics[width= \columnwidth]{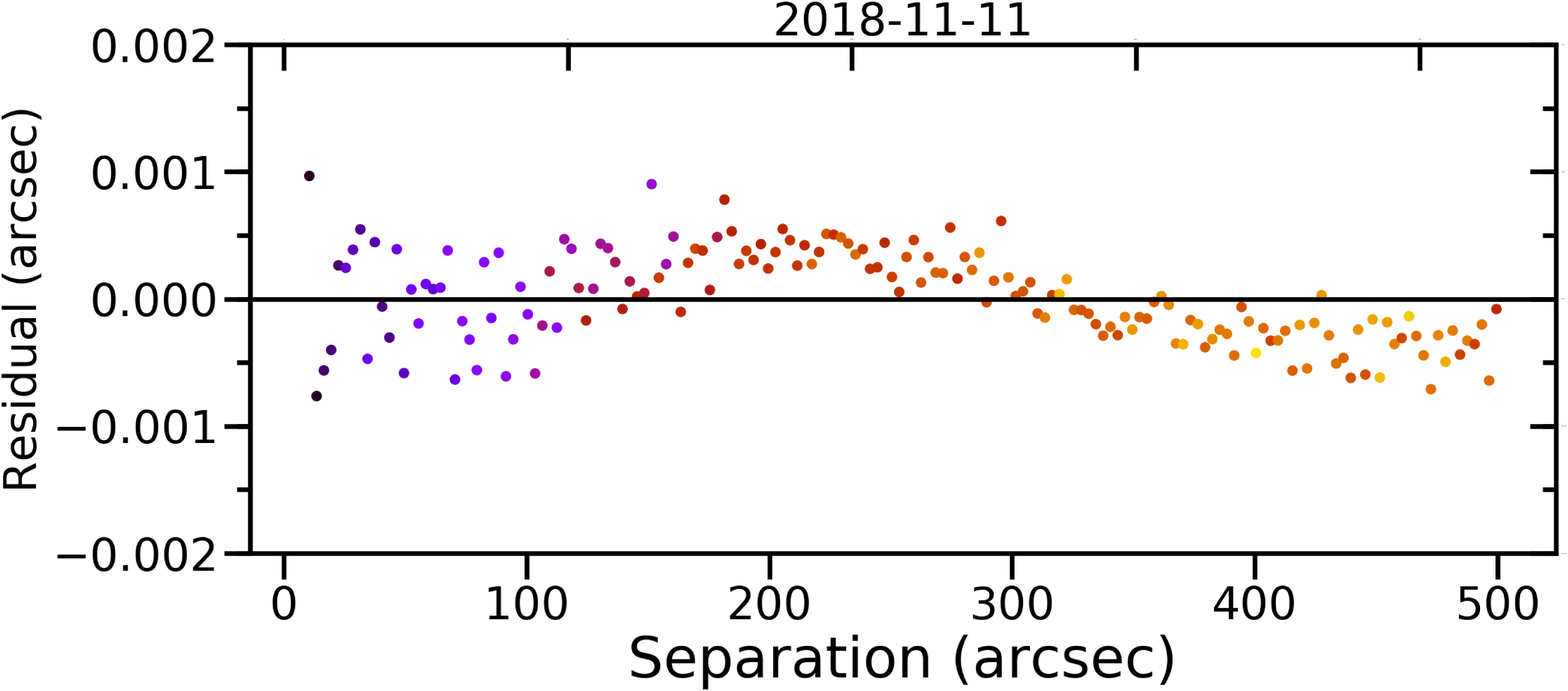}
    \end{minipage}
    \begin{minipage}{0.45\textwidth}
    \vspace{0in}
	\includegraphics[width= \columnwidth]{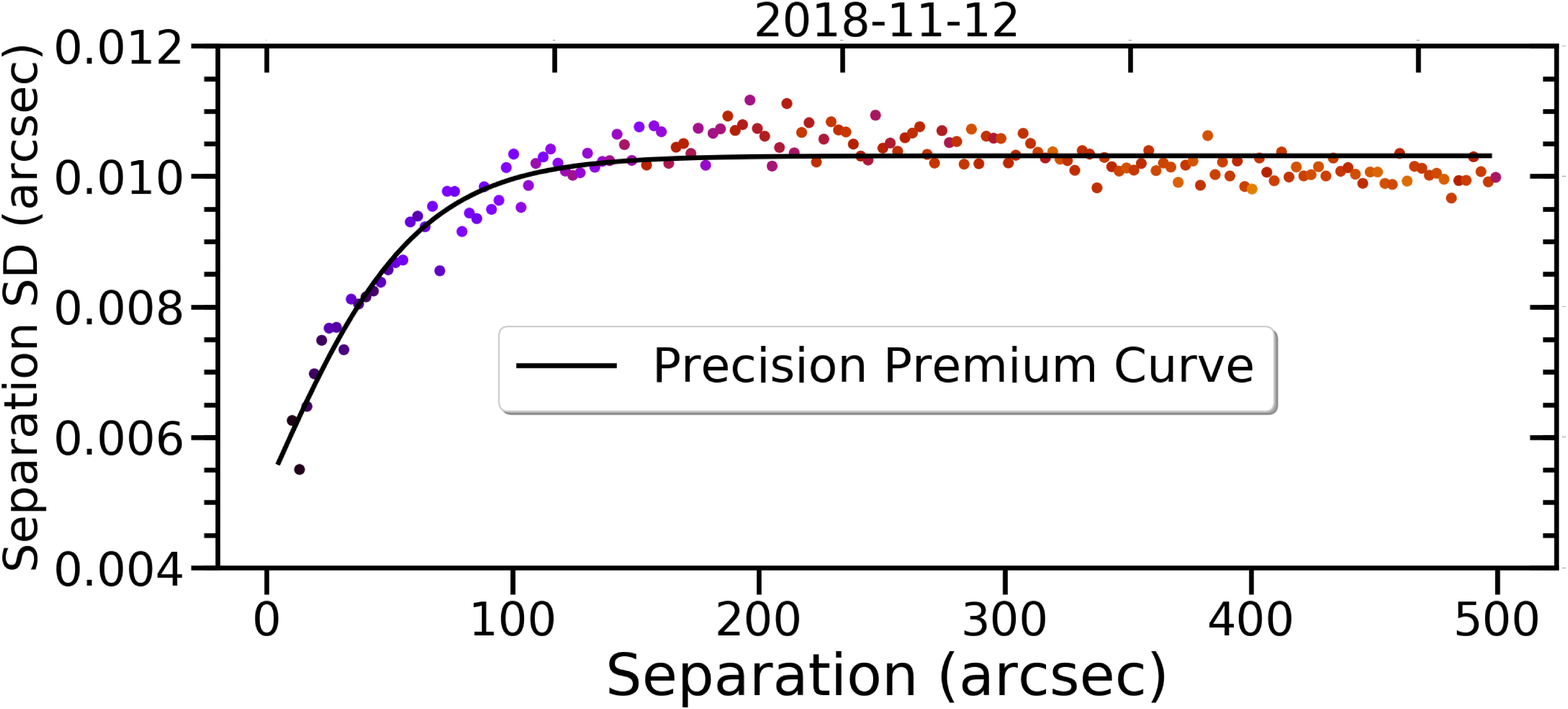}
    \end{minipage}
    \hspace{0.05in}
    \begin{minipage}{0.461\textwidth}
	\includegraphics[width= \columnwidth]{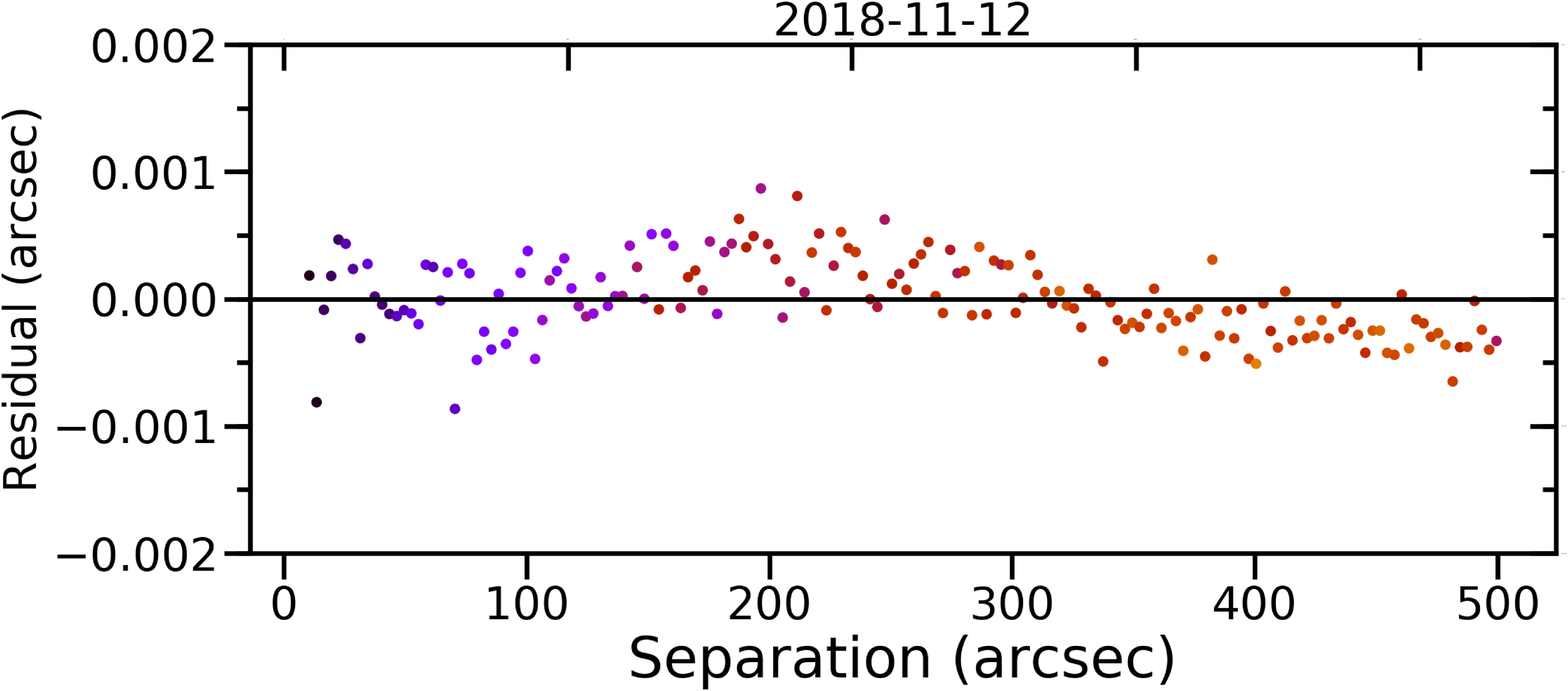}
    \end{minipage}

    \begin{minipage}{0.45\textwidth}
    \vspace{0in}
	\includegraphics[width=\columnwidth]{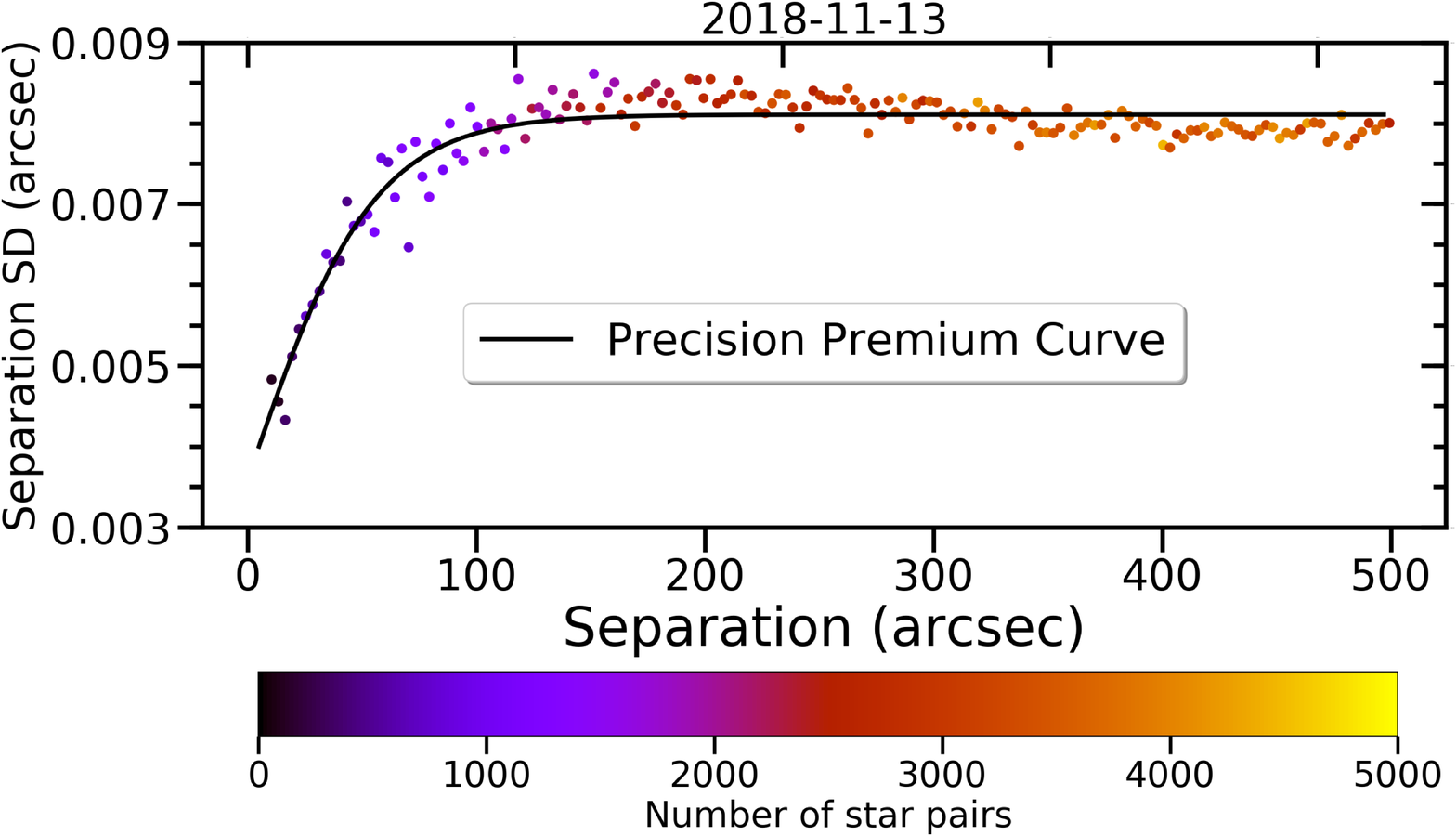}
     \end{minipage}
    \begin{minipage}{0.461\textwidth}
    \vspace{-0.371in}
	\includegraphics[width= \columnwidth]{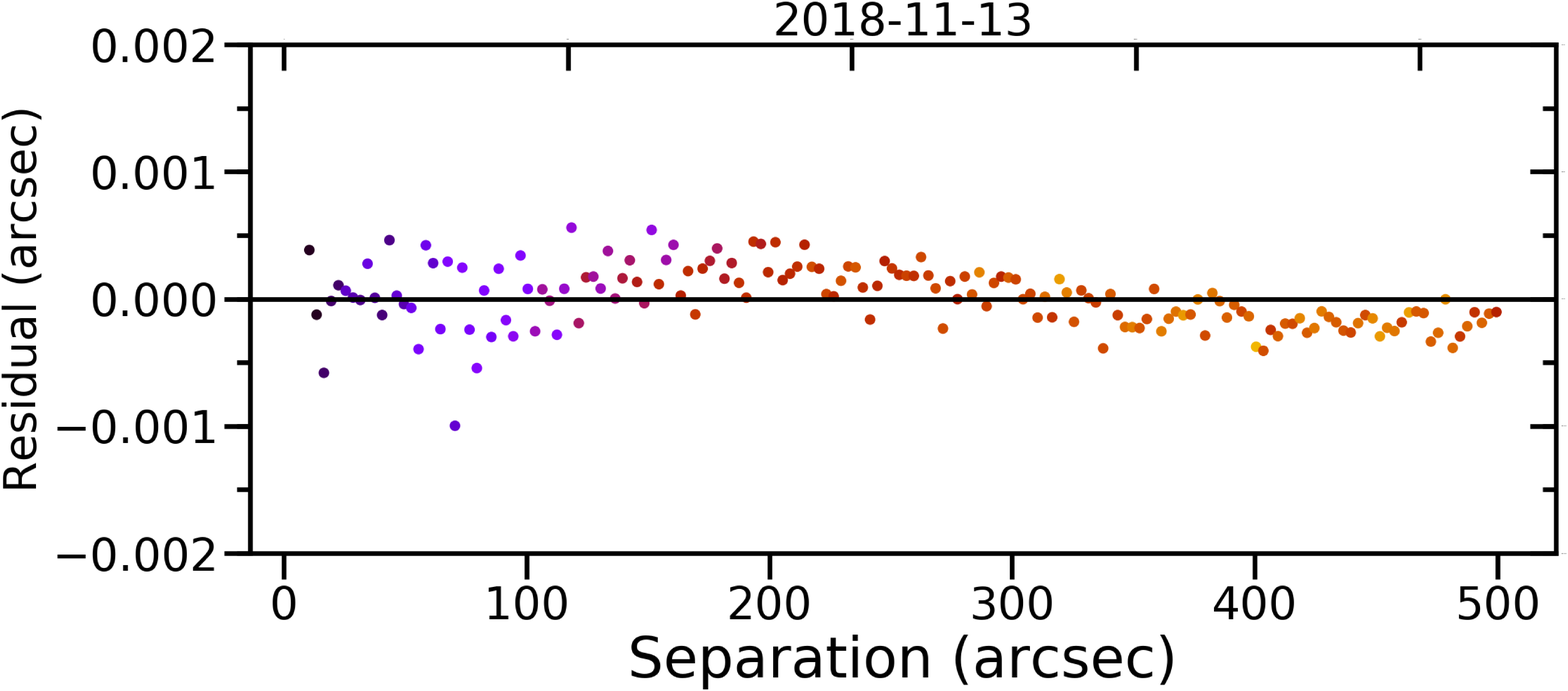}
    \end{minipage}
    \caption{Left panels: statistics of the separation residuals and corresponding PPC of observations listed in Table~\ref{tab:observation}. The horizontal axis is the separation of the star pairs calculated by their standard coordinates and the vertical axis is SD of the separation. The color of the dot represents the number of star pairs used to calculate each SD. Right panels: residuals of the PPC fitting in the corresponding left panel.}
    \label{fig:ppCurve}
\end{figure*}

After the separation and its residual of each star pair in each frame are calculated, the effect of precision premium can be seen clearly from the collection of these separation residuals. Figure~\ref{fig:scatter} shows an example in which each point expresses a separation residual (absolute value) of a star pair in each CCD frame captured on November 13. Faint stars (fainter than 14 Gaia G-mag for observations in this paper) should not be used in this process, because precision premium among them is gradually overwhelmed by their centering errors when their signal-to-noise ratios decrease. We can see from the figure that when the separation of a star pair is less than 100 arcsecs, the residual is obviously smaller as the separation decreases. It is impossible to deliver directly the specific relation between the separation and its measurement precision using the scatter diagram. Therefore, we calculate a standard deviation (SD) in 3 arcsecs bin of separation. Star pairs with too small separation (<5 arcsecs) are unsuitable for centering measurement purpose, so they are not taken into account here. Figure~\ref{fig:ppCurve} shows the statistics of these separation residuals.

Based on the statistics, the PPC can be generated using another sigmoidal curve fitting. The equation of this curve is expressed as
\begin{equation}
\sigma(s)=(B_1-B_2)/(1+e^{s/ds})+B_2,
\end{equation}
where $B_1, B_2$ and $ds$ are the parameters to be solved and $s$ is the separation of star pairs. $B_2$ represents the upper threshold of the curve. As can be seen from the left panels of Figure~\ref{fig:ppCurve}, the sigmoidal curve is suitable for describing precision premium.

\begin{figure*}
	\includegraphics[width= \columnwidth]{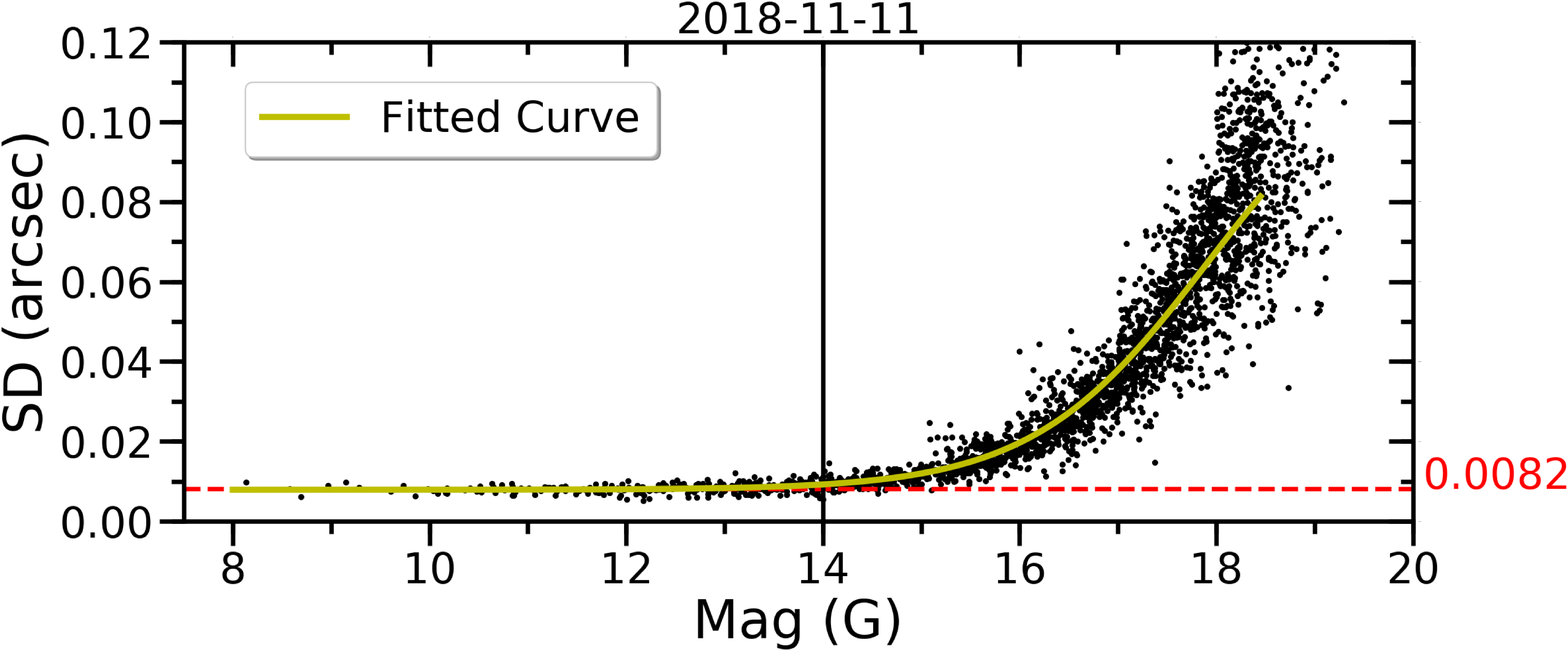}
    \includegraphics[width= 0.94\columnwidth]{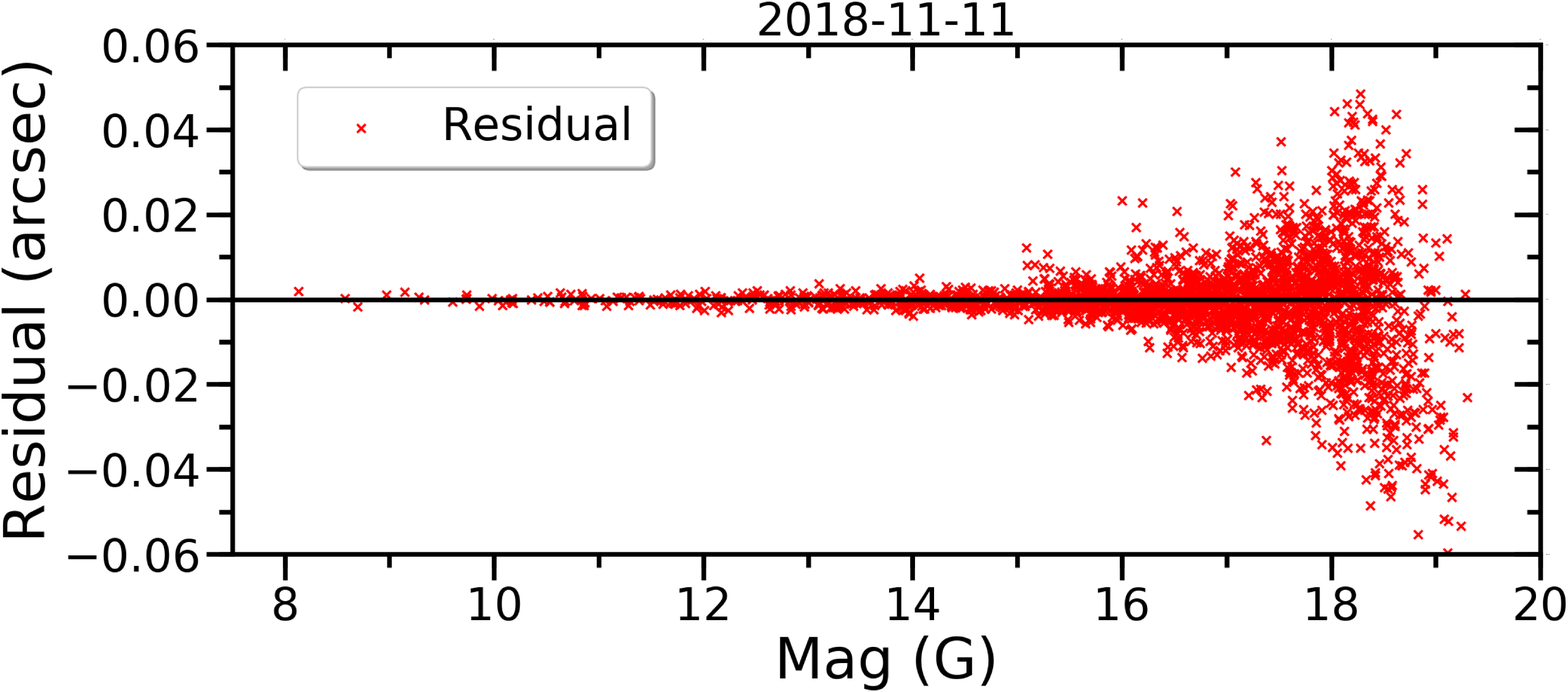}
	\includegraphics[width= \columnwidth]{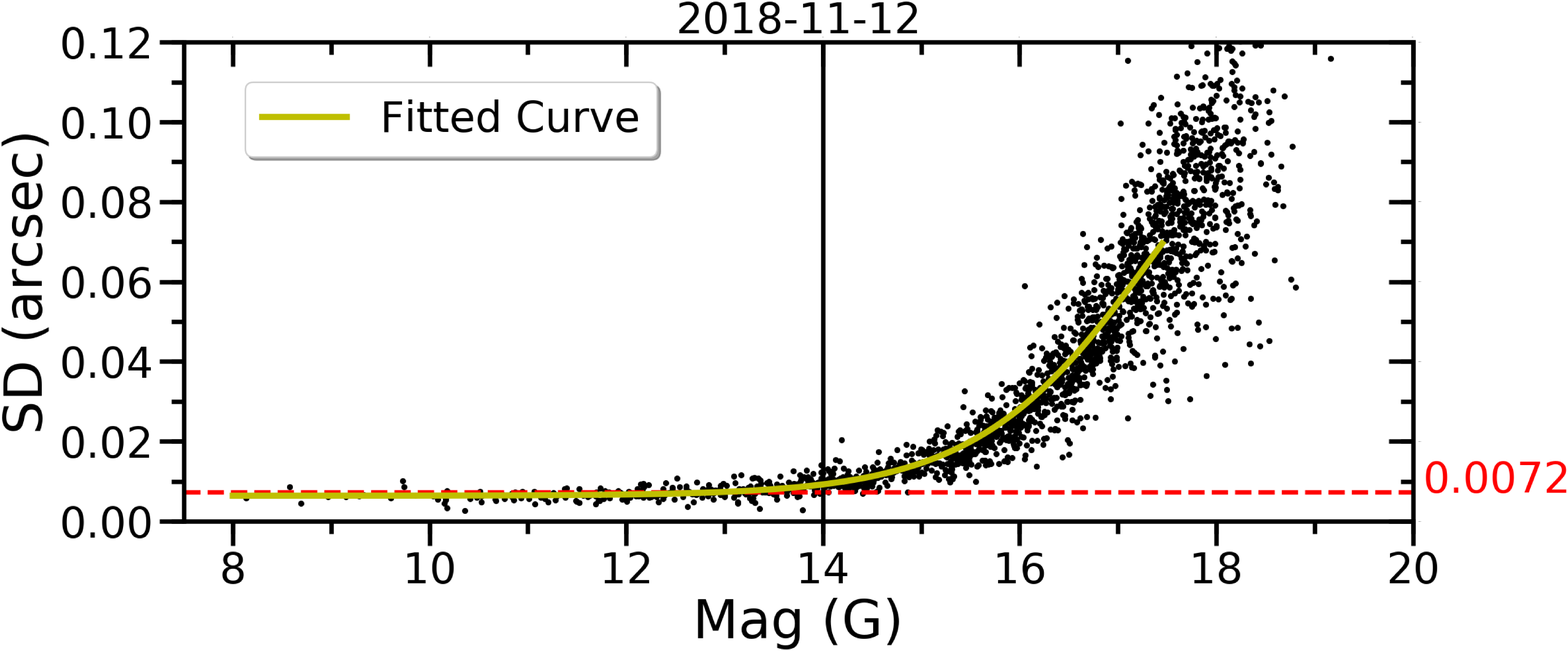}
    \includegraphics[width= 0.94\columnwidth]{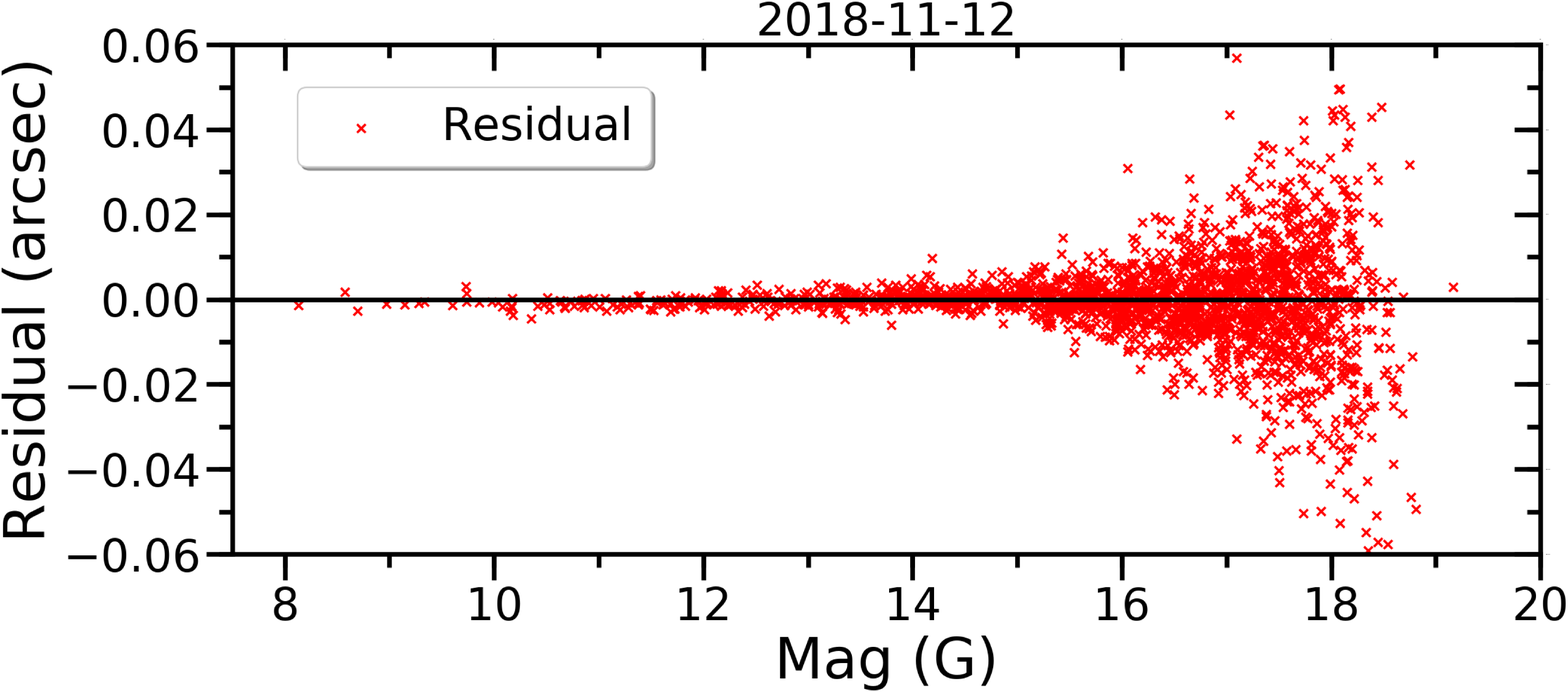}
	\includegraphics[width=\columnwidth]{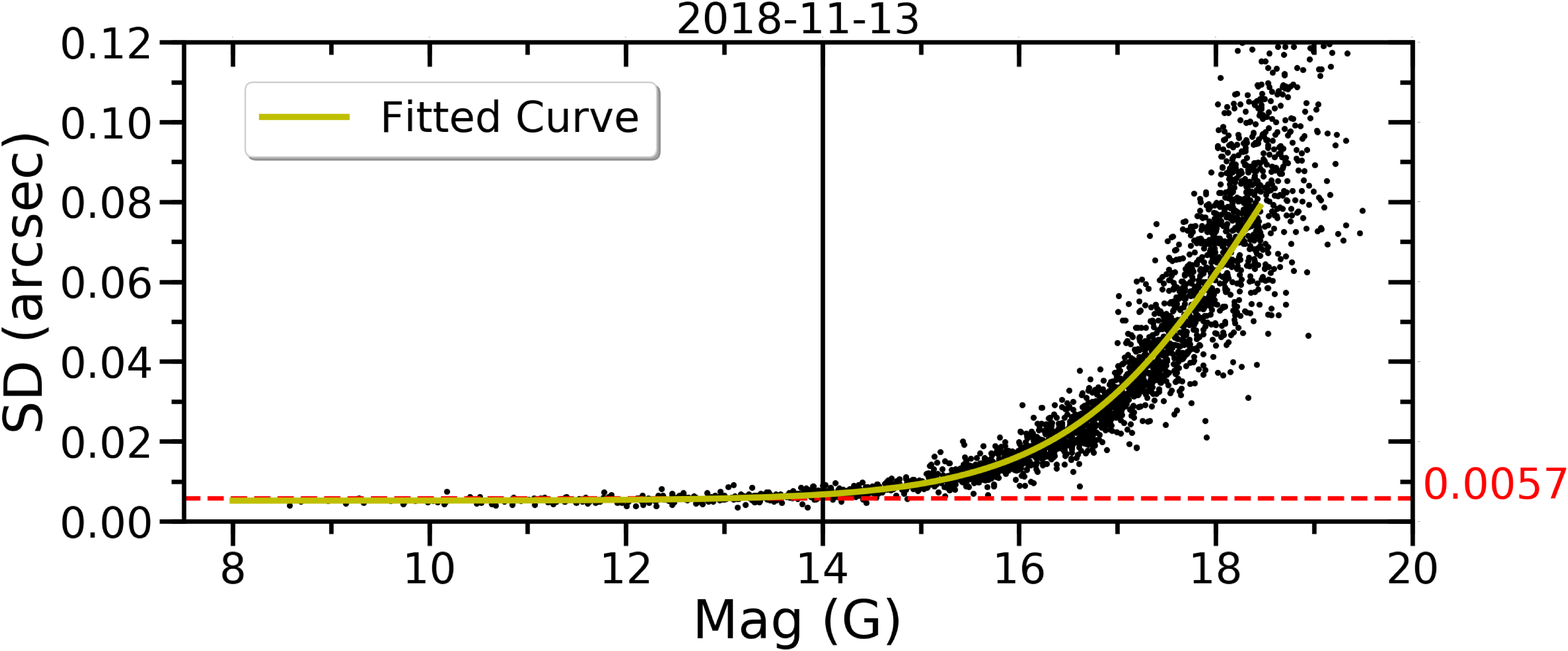}
    \includegraphics[width=0.94 \columnwidth]{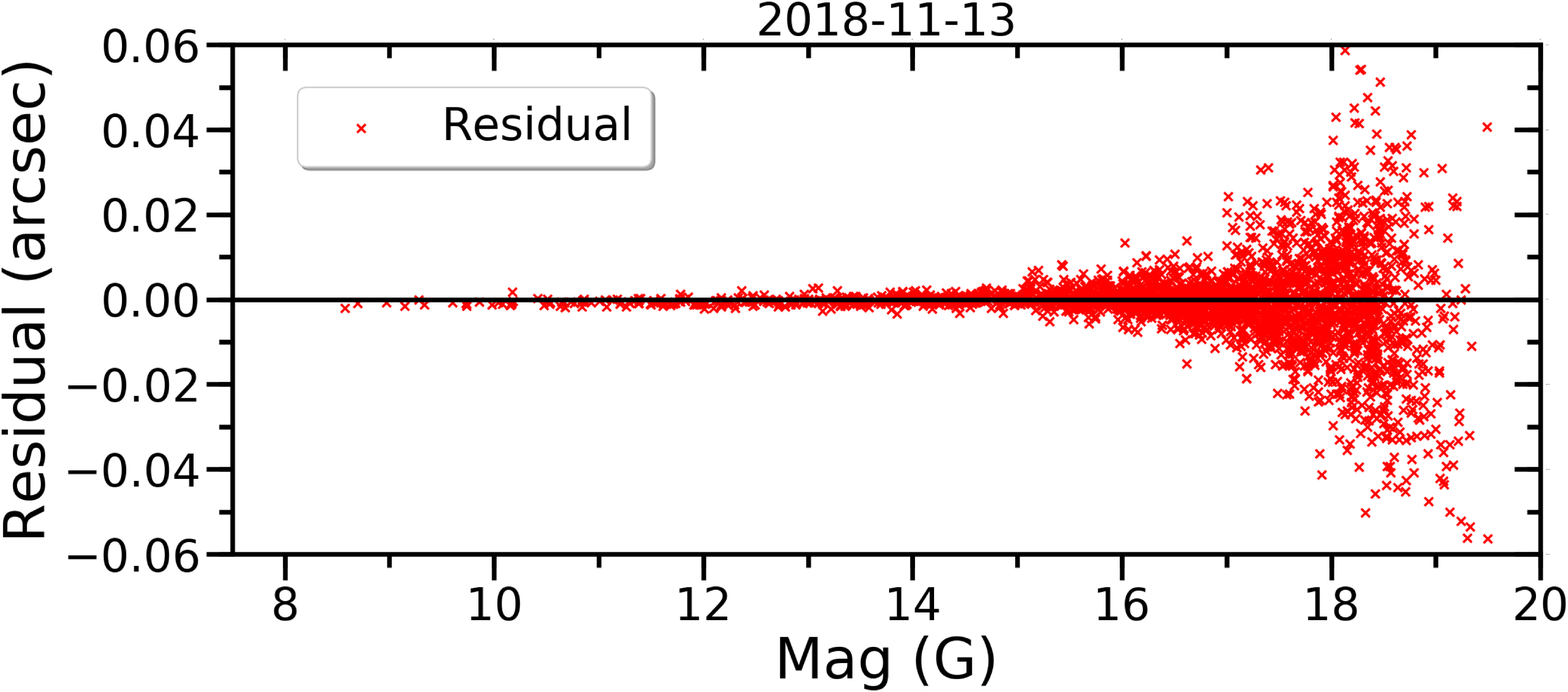}
    \caption{Left panels: reduction of the open cluster observations listed in Table~\ref{tab:observation}. The horizontal axis is Gaia G-mag and the vertical axis is positional standard deviation (SD) calculated by $\scriptstyle{\sigma_E=\sqrt{\sigma_\alpha^2+\sigma_\delta^2}}$. The horizontal dashed line marks the median of the positional SDs for stars brighter than 14 Gaia G-mag, the number on the right is the median. Right Panels: residuals of the curve fitting in the corresponding left panel. }
    \label{fig:mpData}
\end{figure*}

Let the astrometric precision of a star (when its position is solved by a classical plate constant model) be $\sigma_{E}$. Since the single measurement error of a star can be approximated as a Gaussian random variable, the precision of the separation between a star pair whose $(O-C)s$ are independent should satisfy the following equation
\begin{equation}
\label{eq_r}
\sigma_{E}^{}=B_2/\sqrt{2}.
\end{equation}
The median in each panel of Figure~\ref{fig:mpData} is very close to its $\sigma^{}_{E}$ calculated from the parameter $B_2$ (see Table \ref{tab:curveParam2}) of the PPC on the corresponding panel in Figure~\ref{fig:ppCurve}. This indicates that Eq.~(\ref{eq_r}) fits the reduction of the observations well. The theoretical lower limit of the PPC, $(B_1+B_2)/2$, may be limited by centering precision of the star images only. Detailed parameters of the fitted curve in Figure~\ref{fig:mpData} and \ref{fig:ppCurve} are given in Table~\ref{tab:curveParam1} and \ref{tab:curveParam2}, respectively.

\begin{table*}
	\centering
	\caption{Detailed parameters of the fitted curves in the left panels of Figure~\ref{fig:mpData} and the standard deviations of these parameters.}
	\label{tab:curveParam1}
	\begin{tabular}{ccccc} 
		\hline
		Date & $A_1$ & $A_2$ & $m_0 $& $dm$\\
            & (arcsec)&(arcsec)&(mag)&(mag)\\
		\hline
		2018-11-11 & $0.0079\pm0.0006$ &$ 0.179\pm0.019$ & $18.4\pm0.2$ &$0.942\pm0.050$\\
		2018-11-12 & $0.0072\pm0.0006$ &$ 0.123\pm0.007$ & $17.3\pm0.1$ &$0.809\pm0.041$\\
		2018-11-13 & $0.0059\pm0.0005$ &$ 0.183\pm0.020$ & $18.6\pm0.2$ &$0.911\pm0.045$\\
		\hline
	\end{tabular}
\end{table*}

\begin{table*}
	\centering
	\caption{Detailed parameters of the fitted curves in the left panels of Figure~\ref{fig:ppCurve} and the standard deviations of these parameters. The fourth column is calculated by Eq.~(\ref{eq_r}).}
	\label{tab:curveParam2}
	\begin{tabular}{ccccc} 
		\hline
		Date & $B_1$ & $B_2$ & $ds$ & $\sigma_E$\\
            & (arcsec)&(arcsec)&(arcsec)&(arcsec)\\
		\hline
		2018-11-11 & $-0.00247\pm0.00049$ & $0.0120\pm0.00003$& $31.3\pm1.3$ &0.0083\\
		2018-11-12 & $\ \ 0.00006\pm0.00044$ & $0.0103\pm0.00003$& $30.1\pm1.5$& 0.0072\\
		2018-11-13 & $-0.00093\pm0.00037$ & $0.0081\pm0.00002$& $27.5\pm1.3$&0.0057\\
		\hline
	\end{tabular}
\end{table*}
Precision premium can be used to improve the astrometric precision since its characteristics are clarified. The most direct application is in mutual approximation technique \citep{Morgado2016, Morgado2019}. This paper demonstrates that the precision premium exists between any two objects with less than 100 arcsecs separation. And the mutual approximation technique can be adapted to any other objects in relative motion. Moreover, the PPC provides the weights for fitting the observed distance in this technique, so the weighted least squares fitting can be used to improve the result of two objects with fast relative motion.

When the advantages of precision premium are brought into play, the suggestion by \citet{hk2014} that the accuracy of ground-based observations can reach 1 mas is very promising. But for Galilean satellites, the accuracy will be lower because the irregularity of the figure of the satellites will set a limit \citep{hk2014}.

\subsection{An example of precision premium}
\begin{figure}
\begin{center}
	\includegraphics[width= 0.8\columnwidth]{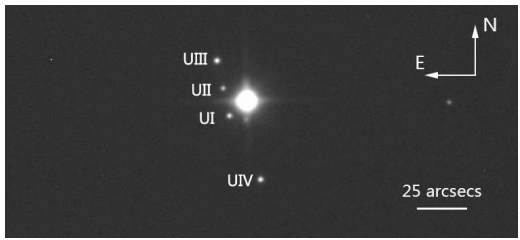}
    \caption{A typical CCD frame (partly) for the Uranian satellites on November 11, 2018.}
    \label{fig:uraobs}
\end{center}
\end{figure}

A good example to show precision premium is the reduction of Uranian satellites' observation captured on November 11. The detailed information of the observations is given in Table~\ref{tab:observation}. In these observations, Uranian satellites have an appropriate separation between each other, as shown in Figure~\ref{fig:uraobs}. Miranda is not considered here since it is faint and affected by the halo of Uranus seriously.

\begin{figure}
  \centering
  \includegraphics[width=\columnwidth]{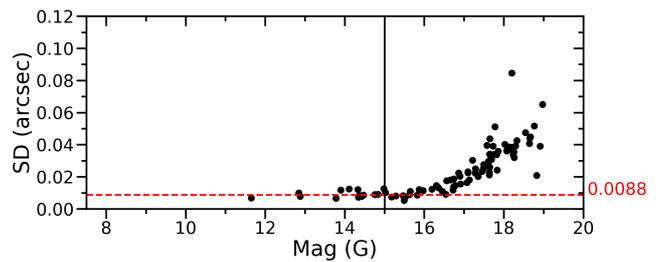}
  \caption{Astrometric precision of the stars in the reduction of Uranian satellites' observations listed in Table~\ref{tab:observation}. The horizontal dashed line marks the median of the positional SDs for stars brighter than 15 Gaia G-mag.}
  \label{fig:uraStarRes}
\end{figure}

The observations are reduced to obtain the (O-C)s (the difference between our astrometry and the ephemeris) of each satellite. The plate model with a second-order polynomial is used in the reduction. The ephemerides of Uranian satellites are retrieved from Institute de M$\rm\acute{e}$chanique C$\rm\acute{e}$leste et de Calcul des $\rm\acute{E}$ph$\rm\acute{e}$m$\rm\acute{e}$rides (IMCCE), that is, the satellite ephemeris by Lainey \citep{Lainey2008, Arlot2016} and planetary ephemeris DE431. Phase effects of the Uranian satellites are not considered in the reduction since their phases change only slightly in one night and we only care about the dispersion of the residuals. Figure~\ref{fig:uraStarRes} shows the astrometric precision of the reference stars. For a well-exposed reference star, the precision is about 6 mas in RA and 7 mas in DE.

\begin{figure}
\begin{center}
	\includegraphics[width=0.99 \columnwidth]{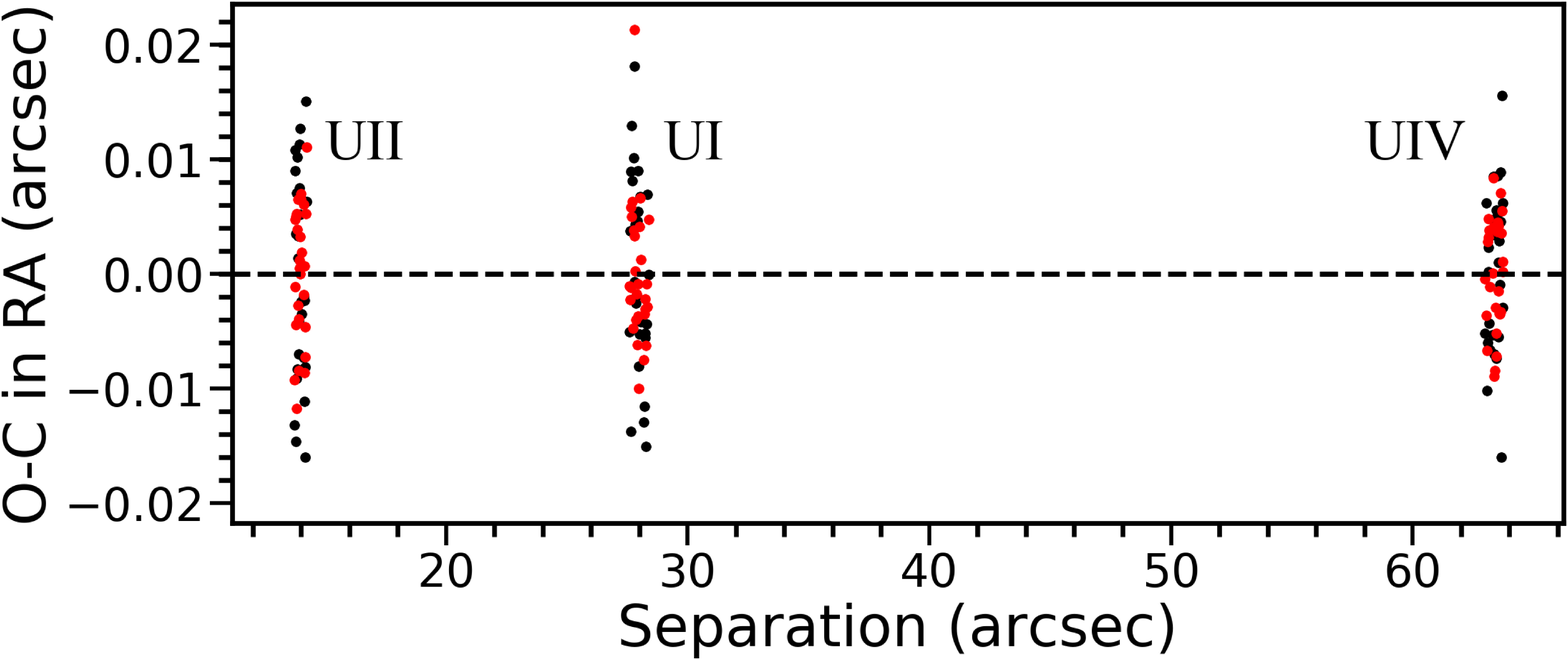}
	\includegraphics[width=0.99 \columnwidth]{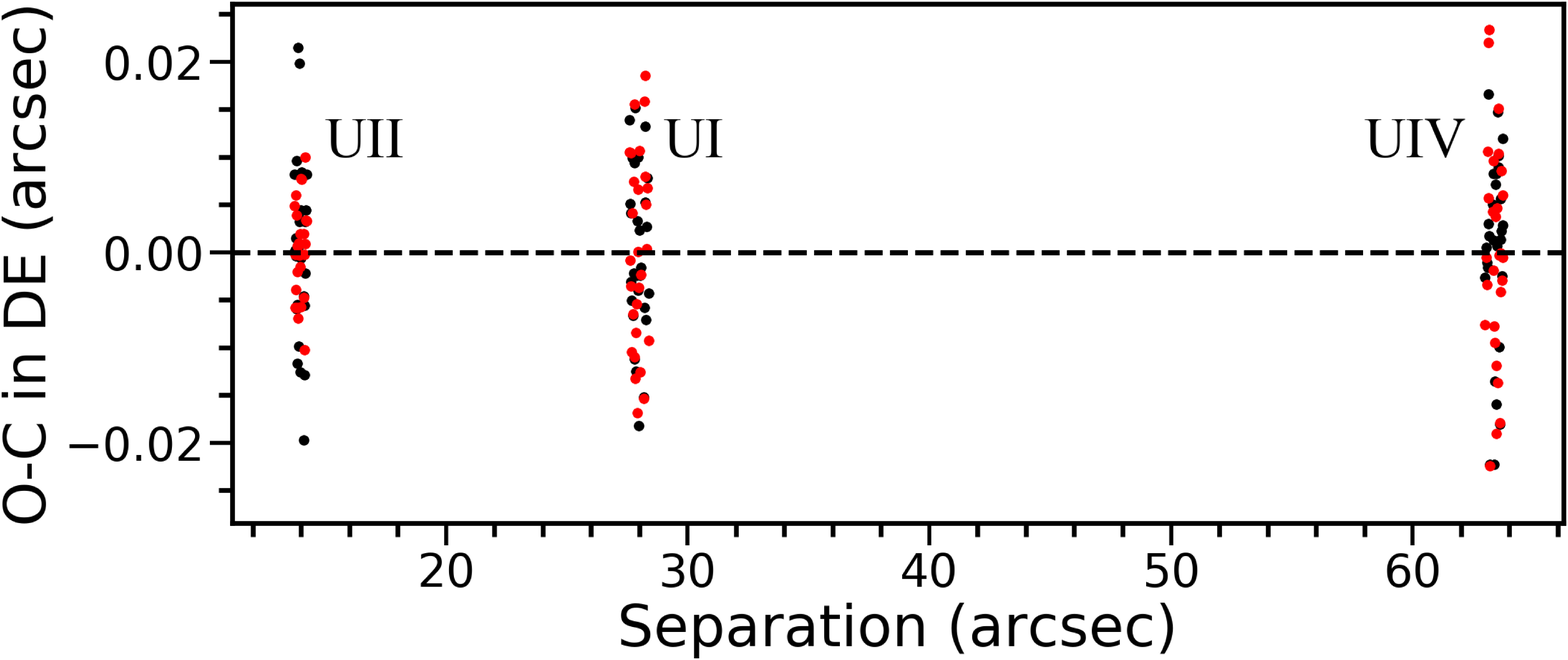}
    \caption{Comparison of the dispersion of the (O-C) residuals before and after precision premium is used. Note that only the standard deviation (SD) of these residuals should be compared here, so the mean (O-C) of each satellite is subtracted. The black dots represent results reduced by classical procedures and the red dots are results with respect to U\uppercase\expandafter{\romannumeral3}. An outlier appears when the precision is improved (see the results of U\uppercase\expandafter{\romannumeral1} in right ascension). }
    \label{fig:uraPpRes}
\end{center}
\end{figure}

The satellite with highest centering precision (brightest and almost unaffected by halo of Uranus), namely U\uppercase\expandafter{\romannumeral3} (Titania) shown in Figure~\ref{fig:uraobs}, is selected as the reference object. Then the positional residuals of other three bright satellites with respect to Titania are calculated and given in Table~\ref{tab:uraRes}. The results are also plotted in Figure~\ref{fig:uraPpRes}, in which the black dots are (O-C) reduced by classical procedures and the red dots are (O-C) with respect to U\uppercase\expandafter{\romannumeral3} (i.e. results that benefit from precision premium). It can be seen from the figure that precision premium for U\uppercase\expandafter{\romannumeral2} (Umbriel) is significant since it is the closest to U\uppercase\expandafter{\romannumeral3} (see Figure~\ref{fig:uraobs}). As mentioned in Sect.~\ref{sec:analysis}, precision premium gradually disappear when the separation between an object and U\uppercase\expandafter{\romannumeral3} increases.

We suggest that if there is any bright object near the target, the bright object should be used as a reference to calculate the residual. In this way, the precision of the target could be improved. The improvement of the precision could be significant if the separation between a target and its reference object is small, which can be seen from the PPC generated using open cluster observations.


\begin{table}
	\centering
	\caption{Statistics of (O-C) residuals for the three satellites of Uranus, which is also showed in Figure~\ref{fig:uraobs}). Column 1 lists the different object. In the 2nd column, \emph{class} denotes the residuals in this row are obtained by the classical reduction and \emph{pp} denotes the results are with respect to U\uppercase\expandafter{\romannumeral3}. The 3rd column is the separation between the object and U\uppercase\expandafter{\romannumeral3} on the observation time, it does not change significantly during the observation period. The following columns list the standard deviation (SD) of the (O-C)s in right ascension and in declination, respectively.}
	\label{tab:uraRes}
	\begin{tabular}{cccrrrr} 
		\hline
		Object & & Separation & SD$_{\alpha}$  &  SD$_{\delta}$\\
        & & (arcsec)& (mas)&(mas)\\
		\hline
		U\uppercase\expandafter{\romannumeral2} & \emph{class} & 14  & 8.8  &9.1\\
		        & \emph{pp} &     & 5.8   &4.8\\
		U\uppercase\expandafter{\romannumeral1} & \emph{class} & 28  & 8.3  &8.7\\
		        & \emph{pp} &     & 6.0  &9.9\\
		U\uppercase\expandafter{\romannumeral4} & \emph{class} & 64 & 7.0 & 10.3\\
		        & \emph{pp} &     & 4.7 & 11.2\\
		\hline
	\end{tabular}
\end{table}

\section{Conclusions}
Three nights of the observations of the open cluster M35 taken with the 1 m telescope at Yunnan Observatory were processed. Statistics of these results demonstrated that precision premium generally exists in the ground-based astrometry and is independent of the relative motion between the objects. The characteristics of precision premium were described using the generated PPCs from practical observations. It is shown that the effective range of precision premium ($\scriptstyle{\lesssim}$100 arcsecs) was stable in three consecutive nights. The upper threshold of the PPC approximates $\sqrt{2}$ times measurement error in a single exposure and the theoretical lower limit of the curve is considered corresponding to the centering precision of the target and its reference object. The PPC demonstrates the potential to achieve a very high precision in conventional astrometry.

\section*{Acknowledgements}

This work was supported by the National Natural
Science Foundation of China (Grant Nos. 11873026, 11703008, 11273014), by
the Joint Research Fund in Astronomy (Grant No. U1431227) under
cooperative agreement between the National Natural Science
Foundation of China (NSFC) and Chinese Academy Sciences (CAS), and
partly by the Fundamental Research Funds for the Central
Universities. The authors would like to thank the chief scientist
Qian S. B. of the 1 m telescope and his working group for their
kindly support and help. This work has made use of data from the European Space Agency (ESA) mission \emph{Gaia} (\url{https://www.cosmos.esa.int/gaia}), processed by the \emph{Gaia} Data Processing and Analysis Consortium (DPAC, \url{https://www.cosmos.esa.int/web/gaia/dpac/consortium}). Funding for the DPAC has been provided by national institutions, in particular the institutions participating in the \emph{Gaia} Multilateral Agreement.





\bsp	
\label{lastpage}
\end{document}